\documentclass[showpacs,showkeys,amsmath,amssymb,12pt]{revtex4}

\usepackage{graphicx}

\begin{document}

\title{Kerr--Sen dilaton-axion black hole lensing in the strong deflection limit}

\author{Galin N. Gyulchev\footnote{E-mail: gyulchev@phys.uni-sofia.bg}, Stoytcho S. Yazadjiev\footnote{E-mail: yazad@phys.uni-sofia.bg} }

\affiliation{Department of Theoretical Physics, Faculty of Physics, Sofia University,\\
5 James Bourchier Boulevard, 1164 Sofia, Bulgaria
}

\date{\today}

\begin{abstract}
In the present work we study numerically quasi-equatorial lensing by the charged, stationary,
axially-symmetric Kerr--Sen dilaton-axion black hole in the strong deflection limit. In this approximation we compute the
magnification and the positions of the relativistic images. The most outstanding effect is that the Kerr--Sen black hole
caustics drift away from the optical axis and shift in clockwise direction with respect to the Kerr caustics.
 The intersections of the critical curves on the equatorial plane as a function of the black hole angular
momentum are found, and it is shown that they decrease with the increase of the parameter $Q^{2}/M$. All of the lensing quantities
are compared to particular cases as Schwarzschild, Kerr and Gibbons--Maeda black holes.
\end{abstract}



\maketitle

\section{Introduction}

One of the consequences of Einstein's General Theory of Relativity is that light rays are deflected by gravity. Although
this discovery was made in the last century, the possibility that there could be such a deflection had been suspected much earlier, by Newton and Laplace, for the first time. The phenomena resulting from the deflection of electromagnetic radiation in a gravitational field are referred to as \emph{gravitational lensing} (GL) and an object causing a detectable deflection is known as a \emph{gravitational lens}. The basic theory of GL was developed by Liebes \cite{Lie}, Refsdal \cite{Ref} and Bourassa, Kantowski \cite{BouKan}. Detailed discussions on GL are the monographs by Schneider, Ehlers, Falco \cite{SEF} and Zakharov \cite{Zak1} and the reviews by Blandford and Narayan \cite{BlaNar}, Refsdal and Surdej \cite{RefSur}, Narayan and Bartelmann \cite{NarBar} and Wambsganss \cite{Wam}.\

Nowadays, gravitational lensing is a rapidly developing area of research, and it has found applications ranging from the
 search of extrasolar planets and compact dark matter to the estimation of the values of the cosmological parameters \cite{S1}. In most of these cases it is possible to assume that the gravitational field is weak, hence the angle of deflection of the light in the field of a spherically symmetric body with mass $M$ can be approximated by: $\hat{\alpha}\approx4GM/uc^{2}$, where $u$ is the impact parameter \cite{SEF}. It is well known that for a Schwarzschild black hole, the deflection angle diverges as $u\rightarrow3\sqrt{3}M$, allowing photons to wind many times around the lens before reaching the observer. In this case, an infinite set of images at both sides of the black hole appears \cite{Group}. In 1959 Darwin \cite{Dar} noticed that photons passing very closely to a black hole may suffer much larger deflections without falling into the event horizon. If an observer is near the line connecting the source and the lens he would detect two infinite series of images very close to the black hole, besides the two classical weak field images. These images are produced by photons, making one or more complete loops around the black hole before reemerging in the observer's direction. These relativistic images could provide a profound test of general relativity in its strong field regime. In this situation, it would be surely possible to distinguish among relativistic theories of gravitation $($e.g. Brans--Dicke, induced gravity, etc$)$. For this reason, it is useful to perform similar investigations within alternative pictures and to evaluate the differences in the results. In this sense, the work of Bekenstein and Sanders, where they considered gravitational lensing by a cluster of galaxies in the frame of scalar tensor theories \cite{BekSan}, may be of interest. In this context Virbhadra, Narasimha and Chitre \cite{VNCh} calculated the Einstein deflection angle for a general static, spherically-symmetric spacetime and constructed circular gravitational lens model characterized with two parameters: Schwarzschild mass and "scalar charge". The authors applied this results to the Einstein-Massless Scalar theory by considering the general static and spherically-symmetric solution, given by Janis, Newman and Winicour and proposed that the present lens model could be able to distinguish observationally a black hole from naked singularity.

The study of the gravitational lensing when light passes very close to a massive body, such as a black hole (a process
also known as gravitational lensing in the \textit{strong deflection limit} (SDL)), has received wide attention in the recent
past. The development of lensing theory in the strong-field regime started with the work of Frittelli, Kling and Newman \cite{FKN} and that of Virbhadra and Ellis \cite{Vir1}. These authors proposed a definition of an exact lens equation without reference to background spacetime, and constructed the exact lens equation explicitly in the case of the Schwarzschild spacetime. They also noticed that the strong-field thin-lens approximation $($without any small angle approximation$)$ describes lensing satisfactorily even at small impact parameters. The Virbhadra--Ellis lens equation takes an intermediary position between the exact lens equation and the quasi-Newtonian approximation. It makes no assumptions to the smallness of bending angles, but it does make approximative assumptions  to the position of the light sources and the observer. In order to the Virbhadra--Ellis lens equation be valid, the spacetime must be asymptotically flat and both the observer and the light sources must be far away from the lens. Moreover, Virbhadra and Ellis \cite{Vir1} studied lensing due to Schwarzschild black hole in an asymptotically flat background by numerical techniques. They defined the relativistic images which are very demagnified unless the observer, the lens and the source are very highly aligned and would serve as a proof of the general relativity if they were ever observed. After this paper Virbhadra and Ellis \cite{Vir2} numerically investigated the lensing by naked singularities. On the other hand, Perlick \cite{Per} has considered lensing in a spherically symmetric and static spacetime, based on the lightlike geodesic equation without approximations. Eiroa, Romero and Torres \cite{ERT} have described Reissner--Nordstr¨om black hole lensing, confirming the appearance of a similar patterns of images. Later on, the formulas given in Ref. \cite{Pet} were used by Petters to calculate relativistic effects on microlensing events. Bhadra has considered the gravitational lensing due to Gibbons--Maeda--Garfinkle--Horowitz--Strominger $($GMGHS$)$ charged black hole \cite{Bha}. In the last two papers, the authors concentrate on the methods developed by Bozza. Bozza \textit{et al.} \cite{Boz1} developed an analytical technique for obtaining the deflection angle in the strong-field situation and showed that the deflection angle diverges logarithmically as light rays approach the photon sphere of a Schwarzschild black hole. In \cite{Boz2} Bozza extended the analytical theory of strong lensing for a general class of static spherically symmetric metrics and showed that the logarithmic divergence of the deflection angle at the photon sphere is a common feature of such spacetimes. Bozza and Mancini \cite{BM} used the strong field limit to study the time delay between different relativistic images, showing that different types of black holes are characterized by different time delay. Whisker \cite{Whi} and Eiroa \cite{Eir1} used the strong field limit approach to investigate the gravitational lensing properties of braneworld black holes, still Eiroa \cite{Eir2} recently considered the gravitational lensing by an Einstein--Born--Infeld black hole. Sarkar and Bhadra have studied the strong gravitational lensing in the Brans--Dicke theory \cite{SB}.

In cases of ordinary lensing, the lens is placed between the source and the observer. The situation when the observer is placed between the source and the lens is called retro-lensing and was studied for the first time by Holtz and Wheeler \cite{HW}. They analyzed only the two stronger images for a black hole placed in the galactic bulge with the Sun as a source and proposed retro-lensing as a new mechanism for searching of black holes. Eiroa and Torres \cite{ET} presented a complete treatment in the strong field limit of gravitational retro-lensing by a static spherically symmetric compact object having a photon sphere. Their results are compared with those corresponding to ordinary lensing in similar strong field situations. Shortly after that, Zakharov \textit{at al.} \cite{ZNDI} discussed a possibility to detect retro-images of the Sun by a Schwarzschild black hole and the formation of mirages near rapidly rotating Kerr black hole horizons. They proposed a procedure to measure masses and rotation parameters through analysis of the form of the mirages.

The SDL method was also applied to wormholes. Tejeiro and Larranaga considered natural wormholes and their astrophysical signatures in various cases. Applying the strong field limit of gravitational lensing theory, they calculated the deflection angle produced by Morris--Thorne wormholes \cite{S2} in asimptotically flat space-times and showed  that wormholes act like convergent lenses. Recently, Nandi, Zhang and Zakharov \cite{NZZ} considered the SDL by traversable Lorentzian wormholes. Their wormhole solutions are considered in the Einstein minimally coupled theory and in the brane world model. The observables in both theories show significant differences from those arising in the Schwarzschild black hole lensing. As a corollary, it is shown that wormholes with zero Keplerian mass also act as ordinary lenses. In cosmological aspect, Mukherjee and Majumdar \cite{MukMaj} considered particale motion and SDL lensing in the metric exterior of a charged dilaton black hole in de Sitter universe. Konoplya \cite{Kon1} analyzed the motion of massless and massive particles around black holes immersed in an asymptotically uniform magnetic field and surrounded by some mechanical structure, which provides the magnetic field, and showed the increases of the distance of minimal approach, the time delay and the bending angle for a ray of light propagating near a black hole. Recently Konoplya \cite{Kon2} considered the Ernst-Schwarzschild solution for a black hole immersed in a uniform magnetic field and estimated corrections to the bending angle and time delay due to presence of weak magnetic fields in galaxies and between galaxies, and also due to influence of strong magnetic field near supermassive black holes.

Several authors have studied gravitational lensing caused by rotating black holes. Motivation for such a study in the SDL phenomenology comes from the fact that previous works on the subject have selected the supermassive black hole hosted by the radio source Sgr A* \cite{Rich, GenSch} as the best candidate for strong field gravitational lensing as it is mentioned in Ref. \cite{BozQuazi}. Unfortunately, in the first work approximations valid only for small deviations from the straight line path are considered, and therefore the proposed methods are not suited for studying relativistic images \cite{Ser,Bra}. In our aspiration to study the phenomenology of the relativistic images in the strong gravitational field of a Kerr black hole, no approximations can be taken and we need to work with the full equations of motion for null rays. There have been numerous articles on the motion of null rays in the gravitational field of a Kerr black hole \cite{CunBar,KVP,Vie1,Vie2}. Besides, some interesting general results have recently been derived through Morse theory \cite{HasPer}.

In Ref. \cite{BozQuazi} Bozza studied the SDL equations for "quasi-equatorial" orbits of photons around a Kerr black hole
and provided analytical expressions for the positions and magnifications of the relativistic images. Recently after that,
 Vazquez and Esteban solved the lens equation far from the equatorial plane for some particular cases \cite{VE}. In their
next work Bozza \textit{et. all} \cite{Boz5} made a considerable step towards the analytical treatment of Kerr lensing, by
restricting the observer in the equatorial plane and solving the general lens equation using perturbative methods for
small values of the black hole spin. The limitation to the equatorial observer is motivated by the fact that the most
important candidate for a black hole, Sgr A*, is likely to have a spin axis perpendicular to the galactic plane, where the
solar system lies, in first approximation. Bozza \textit{et. all} \cite{Boz6} generelized this work by removing
the restriction to observers on the equatorial plane. Amore \textit{at. all} in series of works \cite{Amore} developed the analytical method for calculating the deflection angle of light in a general static and spherically symmetric metrics. Recently, Iyer and Petters \cite{IyePet} developed an analytical perturbation
framework for calculating the bending angle of light rays lensed by a Schwarzschild black hole. Using a perturbation
parameter given in terms of the gravitational radius of the black hole and the light ray's impact parameter, they determined an invariant series for
the strong-deflection bending angle that extends beyond the standard logarithmic deflection term used in the literature.

Presently, the general relativity is not the only viable theory of gravitation. As we have already mentioned, the strong
gravitational lensing could provide profound examination of the space--time around different kinds black holes.
Therefore, following the method of Bozza \cite{BozQuazi} in the present work we wish to study strong gravitational
lensing due to a rotating, charge, dilaton-axion Kerr--Sen black hole \cite{Bla} in the heterotic string theory with the
 aim of investigating the influence of the dilaton and the axion field in the strong deflection limit observations.

The outline of this paper is as follows. The second section contains a derivation of the first order differential system
for null geodesics applied for Kerr-Sen solution. In Sec. III we discuss the lensing equation in the equatorial plane and
 calculate numerically the deflection angle in the strong field limit. In Sec. IV we discuss the quasi-equatorial lensing
 and compute the positions of the caustic points and the magnifications of the images. In Section V the critical curves
and the caustic structure are considered. A discussion of the results is given in Sec. VI.

\section{Rotating Kerr-Sen dilaton-axion black hole and null geodesics}

In 1992 Sen \cite{Sen} was able to find a charged, stationary, axially-symmetric solution \footnote{Alternative derivation of this solution can be seen in \cite{SYazad}} of the field equations by using target space duality, applied to the classical Ker solution. The line element of this solution can be written, in generalized Boyer--Linquist coordinates \cite{BoyLin}, as
\begin{eqnarray}
ds^{2}=&-&\left(1-\frac{2Mr}{\rho^{2}}\right)dt^{2}+\rho^{2}\left(\frac{dr^{2}}{\Delta}+d\theta^{2}\right)-\frac{4Mra\sin^{2}{\theta}}{\rho^{2}}dt{d\phi} \nonumber  \\  &+&\left(r(r+r_{\alpha})+a^{2}+\frac{2Mra^{2}\sin^{2}{\theta}}{\rho^{2}} \right)\sin^{2}{\theta}d\phi^{2} \label{LineElement},
\end{eqnarray}
where function $\Delta$ and $\rho^{2}$ are given by
\begin{eqnarray}
    \Delta &=& r(r+r_{\alpha})-2Mr+a^{2}, \\
    \rho^{2} &=& r(r+r_{\alpha})+a^{2}\cos^{2}{\theta}.
\end{eqnarray}
Here $M$ is the mass of the black hole, $a$ is the specific angular momentum of the black hole, $r_{\alpha}=Q^{2}/M$, $Q$ being the electrical charge of the hole. In the particular case of a static black hole, \textit{i.e.} $a=0$, the solution (\ref{LineElement}) coincides with the GMGHS solution investigated in SDL regime from Bhadra in \cite{Bha}, while in the particular case $r_{\alpha}=0$ reconstructs the Kerr solution.

The Kerr-Sen space is characterized by a spherical event horizon, which is the biggest root of the equation $\Delta=0$ and is equal to

\begin{eqnarray}
  r_{H}=\frac{2M-r_{\alpha}+\sqrt{(2M-r_{\alpha})^{2}-4a^{2}}}{2}.
\end{eqnarray}

Hence follows that $a<\frac{1}{2}\Big{|}1-{\frac{Q^{2}}{2M^{2}}}\Big{|}$. Beyond this critical value of the spin there is no event horizon and causality violations are present in the whole spacetime, with the appearance of a naked singularity. We will restrict to subcritical angular momenta. Radial coordinate, where the Killing vector $\frac{\partial}{\partial{t}}$ is isotropic, \textit{i.e} $g_{tt}=0$, define a surface called ergosphere and is equal to
\begin{eqnarray}
  r_{es}=\frac{2M-r_{\alpha}+\sqrt{(2M-r_{\alpha})^{2}-4a^{2}\cos^{2}\theta}}{2}.
\end{eqnarray}

The null geodesic equations can be derived via the Hamilton--Jacobi equation for metric (1). Following work \cite{Bla}, where Blaga and Blaga examined the geodesics of the Kerr-Sen black hole solution, we can derive the following fourth-order differential system for the null geodesics:
\begin{eqnarray}
    \rho^{2}\dot{r}&=&\pm\sqrt{R(r)}, \label{Sys1} \\
    \rho^{2}\dot{\theta}&=&\pm\sqrt{\Theta(\theta)}, \label{Sys2} \\
    \rho^{2}\dot{\phi}&=&-aE+L_{z}\sin^{-2}\theta+\frac{a}{\Delta}[(r(r+r_{\alpha})+a^{2})E-aL_{z}], \label{Sys3} \\
    \rho^{2}\dot{t}&=&\frac{E(r(r+r_{\alpha})+a^{2})^{2}-2MraL_{z}}{\Delta}-a^{2}E\sin^{2}\theta, \label{Sys4}
\end{eqnarray}
where
\begin{eqnarray}
  R(r)&=&[aL_{z}-(r(r+r_{\alpha})+a^{2})E]^{2}-\Delta[(L_{z}-aE)^{2}+{\cal K}], \label{Func_R} \\
  \Theta(\theta)&=&{\cal K}-\cot^{2}\theta[L^{2}_{z}-\sin^{2}\theta{E}^{2}a^{2}]. \label{Func_Theta}
\end{eqnarray}

 In Eqs. (\ref{Sys1})-(\ref{Sys4}), the dot indicates the derivative with respect to some affine parameter, in Eqs. (\ref{Func_R}) and (\ref{Func_Theta}) ${\cal K}$ is a separation constant of motion, $E=-p_{t}$ is the energy at infinity, $L_{z}=p_{\phi}$ is the angular momentum with respect to the rotation axis of the black hole, while $a=L_{z}/M$ is the angular momentum per unit mass.

The final expression of the lightlike geodesics follows from Eqs. (\ref{Sys1})-(\ref{Sys4}) in terms of integrals
\begin{eqnarray}
  \int^{r}{\frac{dr}{\pm\sqrt{R(r)}}}=\int^{\theta}{\frac{d\theta}{\pm\sqrt{\Theta(\theta)}}},
\end{eqnarray}
\begin{eqnarray}
  \triangle\phi=a\int^{r}{\frac{[(r(r+r_{\alpha})+a^{2})E-aL_{z}]}{\pm\Delta\sqrt{R(r)}}}dr+
                \int^{\theta}\frac{[L_{z}\sin^{-2}\theta-aE]}{\pm\sqrt{\Theta(\theta)}}d\theta.
\end{eqnarray}

The sign of $\sqrt{R(r)}$ and $\sqrt{\Theta(\theta)}$ is positive when the lower integration limit is smaller than the upper limit, and negative otherwise.

We describe the light ray trajectory with the assumption that at infinity, where there is no gravitational field, it is a straight line. In this treatment we can identify the approximate light ray with three parameters. The first of them,  the inclination $\psi_{o}$, is the angle that the incoming light ray forms with the equatorial plane, defined by $\theta=\pi/2$. The projection of the trajectory in the equatorial plane has an impact parameter $u$.

The distance between that point of the projection which is closer to the black hole and the trajectory itself we will denote by $h$. For the observer with coordinates ($r_{o}$, $\theta_{o}$) in the Boyer--Lindquist system we can define two celestial coordinates $\xi_{1}$, $\xi_{2}$ for the images. The coordinate $\xi_{1}$ describes the observable distance of the image with respect to the symmetry axis in direction normal to ray of sight, while the coordinate $\xi_{2}$ represents the observable distance from the image to the source projection in the equatorial plane in the direction orthogonal to the ray of sight. In this formulation using Eqs. (\ref{Sys1}), (\ref{Sys2}) and (\ref{Sys3}) and further allowing $E=1$ and $r_{o}\rightarrow\infty$ we get
\begin{eqnarray}
  \xi_{1}=r_{o}^{2}\sin{\theta_{o}}\frac{d\phi}{dr}\bigg|_{r_{o}\rightarrow\infty}&=&{J}\sin^{-1}{\theta_{o}},\\
  \xi_{2}=r_{o}^{2}\frac{d\theta}{dr}\bigg|_{r_{o}\rightarrow\infty}&=&h\sin{\theta_{o}}.
\end{eqnarray}

Taking under consideration the fact that $\theta_{o}=\pi/2-\psi_{o}$ and setting $\xi_{1}=u$, it is possible to relate the constants of motion $J$ and $\cal{K}$ to the initial conditions characterizing the light ray:
\begin{eqnarray}
  &&J=u\cos{\psi_{o}}, \label{IniCon1} \\
  &&{\cal K}=h^{2}\cos^{2}{\psi_{o}}+(u^{2}-a^{2})\sin^{2}{\psi_{o}}. \label{IniCon2}
\end{eqnarray}

\section{LENSING IN THE EQUATORIAL PLANE}

\subsection{Equation of the gravitational lens in the equatorial plane}

Let us consider the case when both the observer and the source lie in the equatorial plane of the Kerr-Sen black hole and the whole trajectory of the photon is limited on the same plane. If we set the black hole in the origin, then the angle between the direction of the source and the optical axis will be denoted by $\gamma$. $\gamma\simeq0$ is the case of almost perfect alignment. As it is well known (see Ref. \cite{BozQuazi}) the caustic points can be expressed with the strong field limit coefficients and can be estimated numerically as a function of the black hole spin. From the lensing geometry we can write the relation:
\begin{eqnarray}
  \gamma &=& -\alpha(\theta)+\theta+\bar{\theta} \mod 2\pi,
\end{eqnarray}
where according to this, when $u\ll (D_{LS}$, $D_{OL})$ one can write
\begin{eqnarray}
  \bar{\theta} &\simeq& \frac{u}{D_{LS}} \simeq \frac{D_{OL}}{D_{LS}}\theta, \end{eqnarray}
where $\bar{\theta}$ is the impact angle of the source, $D_{LS}$ is the lens--source distance and $D_{OL}$ is the observer--lens distance.

The lens equation in the equatorial plane is then
\begin{eqnarray}
  \gamma &=& \frac{D_{OL}+D_{LS}}{D_{LS}}\theta-\alpha(\theta) \mod 2\pi. \label{LensEq}\end{eqnarray}
The angle $\gamma$ can take any value in the interval $[-\pi$, $\pi]$. $\alpha(\theta)$ is the deflection angle on the looping light ray and it depends directly (through the impact parameter $u$) on the curvature of the space-time geometry of the rotating black hole. As we shall see below, after calculation of the $\alpha(\theta)$ for the Kerr--Sen black hole in the strong field limit we can solve the equatorial lens equation in order to derive the relativistic images.

\subsection{Deflection angle in the equatorial plane}

Let us start with the conditions $\theta=\pi/2$ and $h=\psi_{o}=0$, which set the light ray on the equatorial plane. In this case, if we substitute $x=r/2M$ to a new radial coordinate and measure all distances in units $2M=1$ we obtain the reduced metric in the form
\begin{eqnarray}
  ds^{2} &=& -A(x)dt^{2}+B(x)dx^{2}+C(x)d\phi^{2}-D(x)dt{d}\phi. \label{RMet}
\end{eqnarray}
The metric coefficients are:
\begin{eqnarray}
  A(x) &=& \frac{x+x_{\alpha}-1}{x+x_{\alpha}}, \\
  B(x) &=& \frac{x(x+x_{\alpha})}{x(x+x_{\alpha})-x+a^{2}}, \\
  C(x) &=& x(x+x_{\alpha})+a^{2}+\frac{a^{2}}{x+x_{\alpha}}, \\
  D(x) &=& \frac{2a}{x+x_{\alpha}},
\end{eqnarray}
where $x_{\alpha}=Q^{2}/2M^{2}$.

We express the derivative $\dot{t}$ and $\dot{\phi}$ in terms of the metric coefficients $A(x)$, $C(x)$ and $D(x)$. Using the equations for the constants of motions $E=-p_{t}=-g_{t\mu}\dot{x}^{\mu}$ and $J=p_{\phi}=g_{\phi\mu}\dot{x}^{\mu}$ written for metric (21), we obtain
\begin{eqnarray}
  \dot{\phi} &=& 2\left[\frac{D+2AJ}{4AC+D^{2}}\right], \label{Phidot} \\
  \dot{t} &=& 2\left[\frac{2C-DJ}{4AC+D^{2}}\right]. \label{Tdot}
\end{eqnarray}
Moreover, the Lagrangian $\cal{L}$=$\frac{1}{2}g_{\mu\nu}\dot{x}^{\mu}\dot{x}^{\nu}$ is another constant of motion, which vanishes for null geodesics and can be used to express $\dot{x}$:
\begin{eqnarray}
 \dot{x} &=& \pm2\sqrt{\frac{C-J(D+AJ)}{B(4AC+D^{2})}}. \label{Xdot}
\end{eqnarray}
\begin{figure}
  \includegraphics[width=8cm]{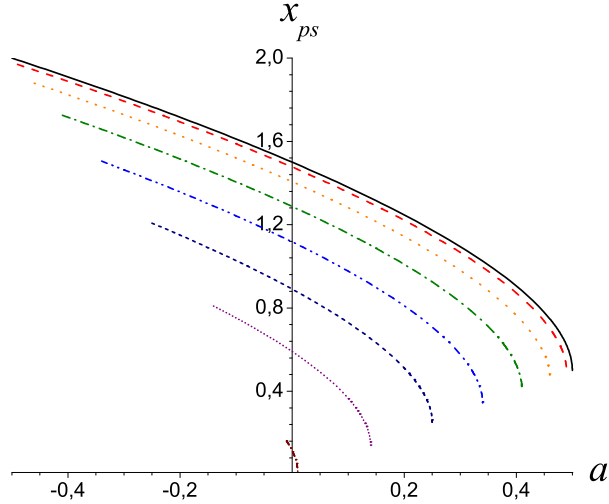}\\
  \caption{The radius of the photon sphere as a function of the Kerr-Sen black hole angular momenta for $x_{\alpha}$ equal to 0 (black, solid line), 0.02, 0.08, 0.18, 0.32, 0.5, 0.72, 0.98 (wine, short, dashed, dotted line).}\label{PhotSp}
\end{figure}
Since in the case under consideration $h=\psi_{o}=0$, immediately from Eq. (\ref{IniCon1}) follows that the angular momentum $J$ is equal to the impact parameter $u$. Evaluating the Lagrangian at the minimum distance of approach of the light ray, where $\dot{x}=0$, with the help of Eqs. (\ref{Phidot}), (\ref{Tdot}) we find the precise relation between $J$ and the closest approach distance $x_{0}$
\begin{eqnarray}
 J=u&=&\frac{-D_{0}+\sqrt{4A_{0}C_{0}+D_{0}^{2}}}{2A_{0}}  \label{ImPar}  \\
    &=&\frac{-a+(x_{0}+x_{\alpha}) \sqrt{x_{0}^{2}+x_{0}(x_{\alpha}-1)+a^{2}}}{x_{0}+x_{\alpha}-1},
\end{eqnarray}
where all metric function are evaluated at $x=x_{0}$. The sign before the square root is chosen to be positive when the light ray is winding counterclockwise. For $a>0$ the black hole rotates counterclockwise, while for $a<0$ the black hole and the photons rotate in converse direction.

A starting point of the strong field limit expansion is the photon sphere, which has been defined by Virbhadra and Ellis \cite{Vir1} and has subsequently been investigated by Claudel, Virbhadra and Ellis \cite{PSphere}. Combining the null geodesics equations and Eqs. (\ref{Phidot})-(\ref{ImPar}) we obtain the photon sphere equation for reduced stationary, axially-symmetric metric (\ref{RMet})
\begin{eqnarray}
   A_{0}C_{0}^{\prime}-A_{0}^{\prime}C_{0}+J(A_{0}^{\prime}D_{0}-A_{0}D_{0}^{\prime})&=&0, \end{eqnarray}
for which we require to admit at least one positive solution. The real root external to the horizon of this equation defines the radius of the photon sphere $x_{ps}=r_{ps}/2M$. For the Kerr--Sen metric the photon sphere equation takes the form
\begin{eqnarray}
  4x_{0}^{4}+12(x_{\alpha}-1)x_{0}^{3} + (13x_{\alpha}^{2}-22x_{\alpha}+9)x_{0}^{2} &+& 2(3x_{\alpha}^{3}-6x_{\alpha}^{2}+3x_{\alpha}-4a^{2})x_{0} \nonumber \\ &+&x_{\alpha}^{4}-2x_{\alpha}^{3}+x_{\alpha}^{2}-4a^{2}x_{\alpha}=0. \end{eqnarray}
The radius of the photon sphere is computed numerically and is plotted in Fig. 1 for different values of $x_{\alpha}$. For all photons, a critical value of the angular momenta $a_{cr}$ exists at which the photons fall on the ergosphere $x_{es}=r_{es}/2M$. In the equatorial case, for the radius of the ergosphere we have $x_{es}=1-x_{\alpha}$. When the photon sphere coincides with the ergosphere (\textit{i.e} $x_{ps}\equiv1-x_{\alpha}$), we can calculate the critical value $a_{cr}$ via the equation of the photon sphere. The values of $a_{cr}$ for different $x_{\alpha}$ are

\begin{center}
  \begin{tabular}{  l | c  c  c  c  c  c  c  r }
    \hline\hline
     $x_{\alpha}$
      & 0  &  0.02  &  0.08  &  0.18  &  0.32  &  0.50  &  0.72  &  0.98  \\
     $a_{cr}$ & 0.354  & 0.348  &  0.332  &  0.304  &  0.262  &  0.204  &  0.124  &  0.01  \\ \hline\hline
   \end{tabular}
\end{center}
When $a\rightarrow\frac{1}{2}\big|1-x_{\alpha}\big|$ then $x_{ps}\rightarrow{x}_{H}$, \textit{i.e} the photon sphere coincides with the horizon in the limit of extremal Kerr--Sen black hole.

Having the expression (26) and (28) we find the azimuthal shift as a function of the distance
\begin{eqnarray}
  \frac{d\phi}{dx} &=& \pm\frac{\sqrt{B|A_{0}|}(D+2JA)}{\sqrt{C}\sqrt{4AC+D^{2}}\sqrt{sgn(A_{0})[A_{0}-A\frac{C_{0}}{C}+\frac{J}{C}(AD_{0}-A_{0}D)]}},
\end{eqnarray}
where the fact that matric coefficient $A_{0}$ is negative under the ergosphere is taken under consideration.

From the symmetry in the phase of approach and departure we can write the whole deflection angle via integration of Eq. (31) from $x_{0}$ to infinity
\begin{eqnarray}
  &&\alpha(x_{0}) = \phi_{f}(x_{0})-\pi,\\
  &&\phi_{f}(x_{0}) = 2\int_{x_{0}}^{\infty}\frac{d\phi}{dx}dx. \label{phi_f_x_0}
\end{eqnarray}

$\phi_{f}(x_{0})$ is the total azimuthal angle. With the decreases of the distance of closest approach $x_{0}$ the deflection angle
increases, and for a certain value of $x_{0}$ the deflection angle becomes $2\pi$, so that the light makes a complete
loop around the black hole. Let $x_{0}$ decreases further, then the light ray will wind several times around the lens
before reaching the observer and finally when $x_{0}$ becomes equal to the radius of the photon sphere $x_{ps}$ the
deflection angle will become unboundedly large and the photon will be captured by the lens object.
Besides, if the photons are winding in the same direction of rotation as the black hole, a much smaller photon
spheres exists for them in comparison to the photon spheres of conversely winding photons. The retrograde photons may be captured more
easily from the black hole than the direct photons.

We can find the behaviour of the deflection angle very close to the photon sphere following the evaluation technique for the integral (\ref{phi_f_x_0}) developed by Bozza \cite{Boz2}.
The divergent integral is first splitted into two parts one of which $\phi_{f}^{D}(x_{0})$ contains the divergence and the other $\phi_{f}^{R}(x_{0})$ is  regular.
Both pieces are expanded around $x_{0}=x_{ps}$ and with sufficiently large accuracy are approximated with the leading terms.
At first, we express the integrand of (\ref{phi_f_x_0}) as a function of two new variables $y$ and $z$ which are defined by
\begin{eqnarray}
   y &=& A(x),\\
   z &=& \frac{y-y_{0}}{1-y_{0}}.
\end{eqnarray}
where $y_{0}=A_{0}$.

The whole azimuthal angle than takes the form
\begin{eqnarray}
  \phi_{f}(x_{0}) = \int_{0}^{1}R(z,x_{0})f(z,x_{0})dz, \label{AzimEngle}
\end{eqnarray}
where the functions are defined as follows
\begin{eqnarray}
  R(z,x_{0}) &=& 2\frac{(1-y_{0})}{A^{\prime}}\frac{\sqrt{B|A_{0}|}(D+2JA)}{\sqrt{C}\sqrt{4AC+D^{2}}}, \label{FuncR} \\
  f(z,x_{0}) &=&  \frac{1}{\sqrt{sgn(A_{0})[A_{0}-A\frac{C_{0}}{C}+\frac{J}{C}(AD_{0}-A_{0}D)]}}. \label{Funcf}
\end{eqnarray}
All functions in (\ref{FuncR}) and (\ref{Funcf}) without the subscript $''0''$ are evaluated at $ x=A^{-1}[(1-y_{0})z+y_{0}]$.
The function $R(z,x_{0})$ is regular for all values of $z$ and $x_{0}$ but $f(z,x_{0})$ diverges when $z\rightarrow0$,  \textit{i. e.} as one approaches the photon sphere. The integral (\ref{AzimEngle}) is then separated in two parts
\begin{eqnarray}
  \phi_{f}(x_{0}) &=& \phi_{f}^{D}(x_{0})+\phi_{f}^{R}(x_{0}),  \end{eqnarray}
where
\begin{eqnarray}
  \phi_{f}^{D}(x_{0}) &=& \int_{0}^{1}R(0,x_{ps})f_{0}(z,x_{0})\end{eqnarray}
contains the divergence and
\begin{eqnarray}
  \phi_{f}^{R}(x_{0}) &=& \int_{0}^{1}g(z,x_{0})dz\end{eqnarray}
is a regular integral in $z$ and $x_{0}$.
To find the order of divergence of the integrand, we expand the argument of the square root of $f(z,x_{0})$ to second order in $z$ and get the function $f_{0}(z,x_{0})$:
\begin{eqnarray}
  f_{0}(z,x_{0}) = \frac{1}{\sqrt{\alpha{z}+\beta{z^{2}}}}, \end{eqnarray}
where:
\begin{eqnarray}
  \alpha &=&  sgn(A_{0})\frac{(1-y_{0})}{A_{0}^{\prime}C_{0}}[A_{0}C_{0}^{\prime}-A_{0}^{\prime}C_{0}+J(A_{0}^{\prime}D_{0}-A_{0}D_{0}^{\prime})], \\
  \beta &=& sgn(A_{0})            \frac{(1-y_{0})^{2}}{2C_{0}^{2}A_{0}^{\prime3}}\{2C_{0}C_{0}^{\prime}A_{0}^{\prime2}+(C_{0}C_{0}^{\prime\prime}-2C_{0}^{\prime2})y_{0}A_{0}^{\prime}-C_{0}C_{0}^{\prime}y_{0}A^{\prime\prime} \nonumber \\
 &&+J[  A_{0}C_{0}(A_{0}^{\prime\prime}D_{0}^{\prime}-A_{0}^{\prime}D_{0}^{\prime\prime})+2A_{0}^{\prime}C_{0}^{\prime}(A_{0}D_{0}^{\prime}-A_{0}^{\prime}D_{0})]\}.
\end{eqnarray}
The function $g(z,x_{0})$ is the difference between the original integrand and the divergent integrand
\begin{eqnarray}
  g(z,x_{0}) &=& R(z,x_{0})f(z,x_{0})-R(0,x_{ps})f_{0}(z,x_{0}).
\end{eqnarray}
When $x_{0}$ becomes equal to $x_{ps}$ the equation of photon sphere holds. Then, the coefficient $\alpha$ vanishes and the leading term of the divergence in $f_{0}$ is $z^{-1}$. Therefore the integral diverges logarithmically. The coefficient $\beta$ takes the form
\begin{eqnarray}
\beta_{ps}=sgn(A_{ps})\frac{(1-A_{ps})^{2}}{2C_{ps}{A_{ps}^{\prime}}^{2}}[A_{ps}C_{ps}^{\prime\prime}-A_{ps}^{\prime\prime}C_{ps}+J(A_{ps}^{\prime\prime}D_{ps}-A_{ps}D_{ps}^{\prime\prime})]. \end{eqnarray}
\begin{figure}
\includegraphics[width=8cm]{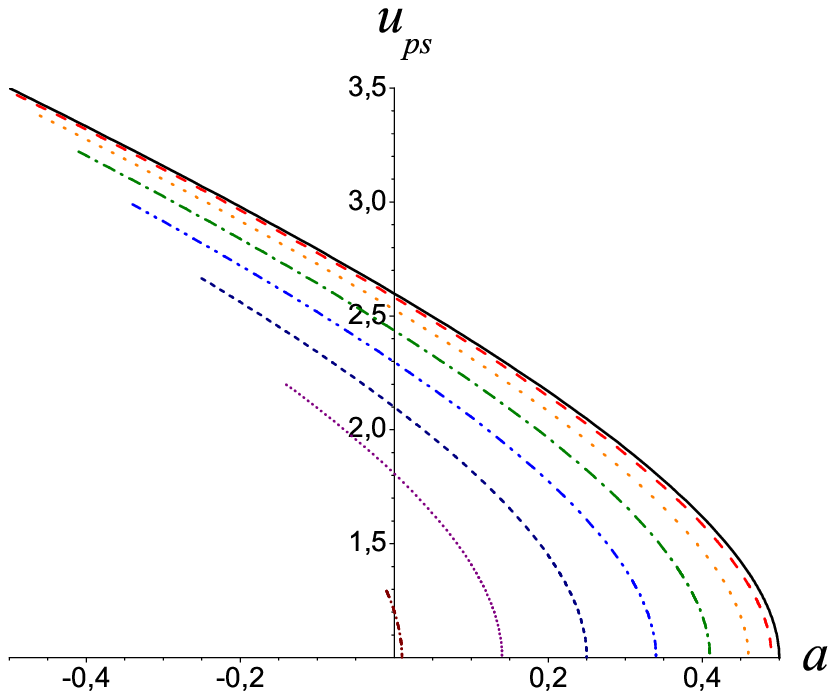}
\includegraphics[width=8cm]{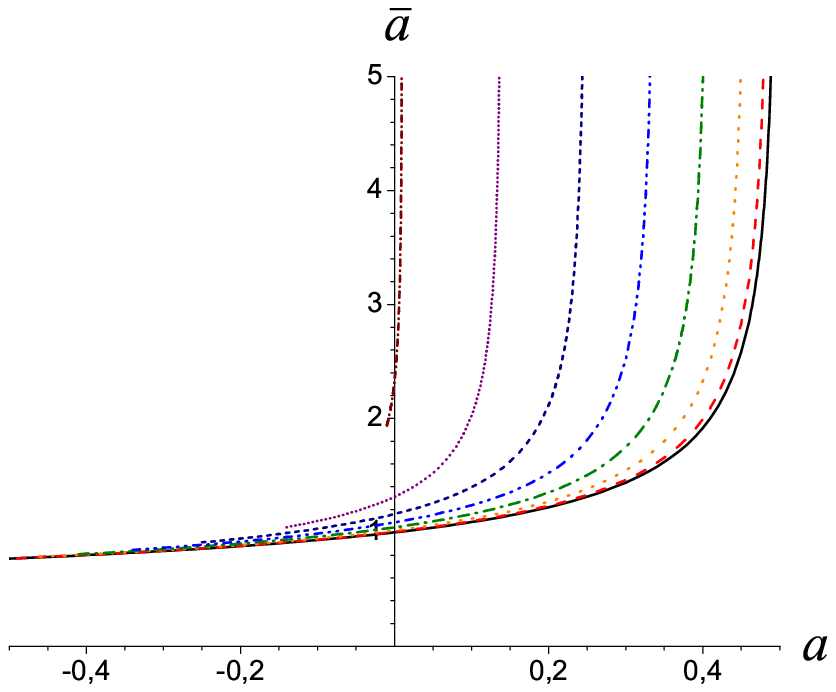}\\ \vspace{0.5cm}
\includegraphics[width=8cm]{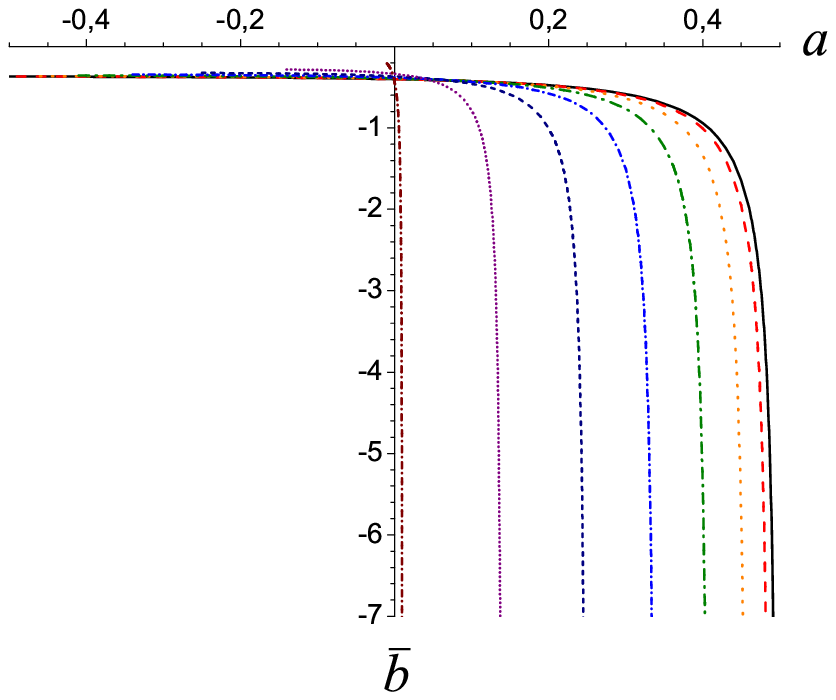}\\
\caption{Strong deflection limit coefficients as a function of the Kerr--Sen black hole angular momenta for $x_{\alpha}$ equal to 0 (black, solid line), 0.02, 0.08, 0.18, 0.32, 0.5, 0.72, 0.98 (wine, short, dashed, dotted line).} \label{KS_J_a_b}
\end{figure}

Close to the divergence, Bozza \cite{Boz2} obtained the analytical expression of the deflection angle expanding the two parts of the original integral (38) around $x_{0}=x_{ps}$ and approximating the leading terms. The result is
\begin{eqnarray}
  \alpha(\theta) &=& -\bar{a}\ln\left( \frac{\theta{D}_{OL}}{u_{ps}}-1 \right) +\bar{b} +O(u-u_{ps}), \label{DefAng}
\end{eqnarray}
where the coefficients $\bar{a}$, $\bar{b}$ and $u_{ps}$ depend on the metric function evaluated numerically at $x_{ps}$.
$D_{OL}$ is the distance between the lens and the observer, and $\theta=u{/}D_{OL}$ is the angular separation between the lens and the image. The strong deflection limit coefficients of the expansion (\ref{DefAng}) are
\begin{eqnarray}
  &&u_{ps}=\frac{-D_{ps}+\sqrt{4A_{ps}C_{ps}+D_{ps}^{2}}}{2A_{ps}},\\
  &&\bar{a} = \frac{R(0,x_{ps})}{2\sqrt{\beta_{ps}}},\\
  &&\bar{b} = -\pi+b_{R}+\bar{a}\ln \left\{\frac{4\beta_{ps}C_{ps}}{u_{ps}|A_{ps}|(D_{ps}+2JA_{ps})}\right\}, \end{eqnarray}
where $b_{R}$ is the regular integral $\phi_{f}^{R}(x_{0})$ evaluated at the point $x_{ps}$ as follows
\begin{eqnarray}
  b_{R} &=& \int_{0}^{1}g(z,x_{ps})dz.\end{eqnarray}
For the same metrics the coefficient $b_{R}$ cannot be obtained analytically. In such cases it must be evaluated numerically.

Fig. 2 shows the strong deflection limit coefficients as functions of $a$. The minimum impact parameter has similar behaviour as the photon sphere $x_{ps}$ and decreases with $a$. $\bar{a}$ grows, while $\bar{b}$ decreases for all values on $x_{\alpha}$. Both coefficients diverge when $a\rightarrow\frac{1}{2}\big{|}1-x_{\alpha}\big{|}$. The divergence of the coefficients of the expansion means that the bending angle in the strong deflection limit (\ref{DefAng}) no longer represents a reliable description in the regime of high $a$.

\subsection{Relativistic images}

Once the deflection angle is known, the positions of the images can be obtained from Eq. (\ref{LensEq}). We consider a generic geometric disposition when the source, the lens and the observer are not aligned.

Let us start with solving the lens equation when $\theta={u}/D_{OL}\ll1$. Then, from the reduced lens equation $\gamma=-\alpha(\theta_{n}^{0})$ mod $2\pi$ and the expression for the deflection angle derived in the strong field limit (\ref{DefAng}), the positions of the n-th relativistic images can be found in first order \cite{BozQuazi}
\begin{eqnarray}
  &&\theta_{n}^{0} = \frac{u_{ps}}{D_{OL}}(1+e_{n}), \\
  &&e_{n}=e^{ \frac{\bar{b}+\gamma-2\pi{n} } {\bar{a}} },
\end{eqnarray}
where n=1, 2, ... indicates the number of winding done by the light ray around the black hole. To build a correction of this solution we can expand $\alpha(\theta)$ around $\theta_{n}^{0}$
\begin{eqnarray}
  \alpha(\theta) &=&   \alpha(\theta_{n}^{0})+\frac{\partial\alpha}{\partial\theta}\bigg|_{\theta_{n}^{0}}(\theta-\theta_n^0)+O(\theta-\theta_n^0)  \nonumber\\
  &\simeq&-\gamma-\frac{\bar{a}D_{OL}}{u_{ps}e_{n}}(\theta-\theta_n^0) \mod 2\pi.
\end{eqnarray}
Replacing this result in the lensing equation (\ref{LensEq}) and neglecting the higher order terms, we find with sufficient precision the position of n-th relativistic images \cite{BozQuazi}
\begin{eqnarray}
  \theta_{n}&\simeq&\theta_{n}^{0}\left(1-\frac{u_{ps}e_{n}(D_{OL}+D_{LS})}{\bar{a}D_{OL}D_{LS}}\right).
\end{eqnarray}

According to the past oriented light ray which starts from the observer and finishes at the source the resulting images stand on the eastern side of the black hole for direct photons ($a>0$) and are described by positive $\gamma$. Retrograde photons ($a<0$) have images on the western side of the black hole and are described by negative values of $\gamma$.

\section{QUASI-EQUATORIAL LENSING}

In order to obtain the caustic structure and to calculate the magnification of the images, following work \cite{BozQuazi}, we need a two dimensional lens equation. Following this plan, we shall restrict for simplicity our consideration to small values of the initial data $\psi_{0}$, and we shall first analyze the ray trajectory close to the equatorial plane for the Kerr--Sen black hole, following the approach of Bozza for quasi-equatorial motion in the strong deflection limit mentioned above. For small declination we have a precession of the photon orbit.

\subsection{Precession of the orbit for small declinations}

We describe the light trajectory with an additional coordinate that is equivalent to the declination $\psi=\frac{\pi}{2}-\theta$. For small declinations and small heights $h$ compared to the projected impact parameter $u$ we have $\psi_{0}\sim{h}/u$. From Eqs. (\ref{IniCon1})-(\ref{IniCon2}) with the accuracy of the first order terms  we get
\begin{eqnarray}
  &&J \simeq u, \\
  &&{\cal K} \simeq h^{2}+\overline{u}^{2}\psi_{0}^{2}, \\
  &&\overline{u} \equiv \sqrt{u^{2}-a^{2}}. \end{eqnarray}

During the motion of the photons we get a simple evolution equation for $\psi$ as a function of the azimuthal angle $\phi$ for the Kerr--Sen black hole \cite{BozQuazi}
\begin{eqnarray}
  \frac{d\psi}{d\phi} &=& \pm\omega(\phi)\sqrt{{\overline{\psi}}^{2}-\psi^{2}}, \label{EqPsiPhi} \end{eqnarray}
where
\begin{eqnarray}
 &&\overline{\psi}=\sqrt{ \frac{h^{2}}{{\overline{u}}^{2}}+\psi_{0}^{2} },  \\
 &&\omega(\phi)=\overline{u}\frac{a^{2}+x(\phi)(x(\phi)+x_{\alpha}-1)}{[a+J(x(\phi)+x_{\alpha}-1)]x(\phi)}. \end{eqnarray}
The solution of Eq. (\ref{EqPsiPhi}), since $x$ depends on $\phi$, is
\begin{eqnarray}
  \psi(\phi)=\overline{\psi}\cos(\overline{\phi}+\phi_{0}) \label{Psi_Phi}, \end{eqnarray}
with
\begin{eqnarray}
  \overline{\phi} &=& \int_{0}^{\phi}\omega(\phi^{\prime})d\phi^{\prime}.\end{eqnarray}

For the photons coming from infinity and returning to infinity according to strong deflection limit the upper integral can be written as follows
\begin{eqnarray}
 \overline{\phi}_{f} &=& \int_{0}^{1} R_{\omega}(z,x_{0})f(z,x_{0})dz,  \end{eqnarray}
where
\begin{eqnarray}
  R_{\omega}(z,x_{0}) &=& \omega(x)R(z,x_{0})\end{eqnarray}
 is a regular function. The functions $R(z,x_{0})$ and $f(z,x_{0})$ are given respectively via Eqs. (\ref{FuncR}) and (\ref{Funcf}).

Then the representation of $\overline{\phi}_{f}$ in the strong field limit is
\begin{figure}
\includegraphics[width=8cm]{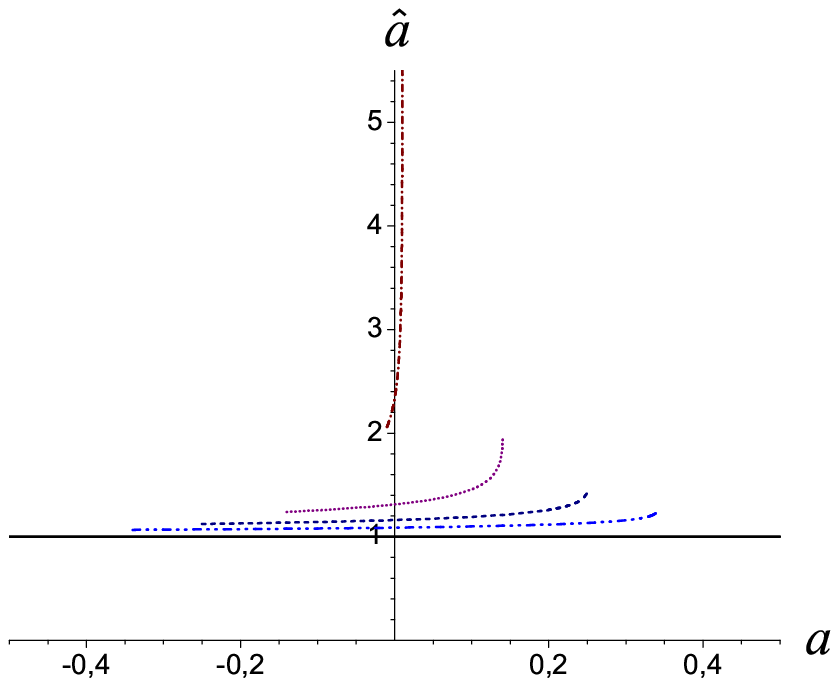}
\includegraphics[width=8cm]{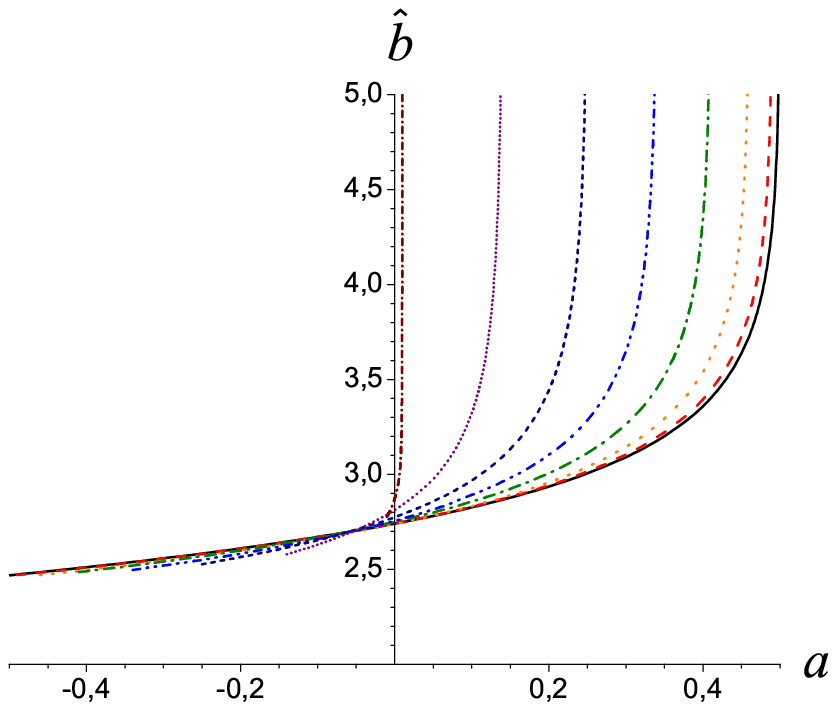}\\
\caption{The coefficients $\hat{a}$ and $\hat{b}$ as a function of a Kerr--Sen black hole angular momentum for $x_{\alpha}$ equal to 0 (black, solid line), 0.02, 0.08, 0.18, 0.32, 0.5, 0.72, 0.98 (wine, short, dashed, dotted line).} \label{KS_a^_b^}\end{figure}
\begin{eqnarray}
  &&\overline{\phi}_{f}=-\hat{a}\ln\left( \frac{\theta{D}_{OL}}{u_{ps}}-1 \right) +\hat{b}, \label{Phi_f} \\
  &&\hat{a}=\frac{R_{\omega}(0,x_{ps})}{2\sqrt{\beta_{ps}}},\\
  &&\hat{b}=\hat{b}_{R}+\hat{a}\ln\left\{ \frac{4\beta_{ps}C_{ps}}{u_{ps}|A_{ps}|(D_{ps}+2JA_{ps})}, \right\}
\end{eqnarray}
where
\begin{eqnarray}
  \hat{b}_{R} = \int_{0}^{1} [R_{\omega}(z,x_{ps})f(z,x_{ps})-R_{\omega}(0,x_{ps})f_{0}(z,x_{ps})] dz. \end{eqnarray}

The quantities $\hat{a}$ and $\hat{b}$ are evaluated numerically in the photon sphere $x_{ps}$.
For the Kerr--Sen metric the coefficient $\hat{a}$ can be calculated analytically and the result is
\begin{eqnarray}
  \hat{a} &=& \left [1+\frac{x_{\alpha}[x_{ps}^{2}+(x_{\alpha}-1)x_{ps}+2a(a-\xi)]}{x_{ps}[x_{ps}^{2}(x_{ps}-1)+2a(a-\xi)+x_{\alpha}[3x_{ps}^2+3(x_{\alpha}-1)x_{ps}+(x_{\alpha}-1)^2]]}\right ]^{\frac{1}{2}},
\end{eqnarray}
where
\begin{eqnarray}
  \xi &=& \sqrt{x_{ps}(x_{ps}+x_{\alpha}-1)+a^{2}}.\end{eqnarray}

In the particular case $x_{\alpha}=0$, the Kerr solution is recovered because of which the coefficient $\hat{a}$ is equal to 1 for all values of the black hole spin. The behaviour of $\hat{a}$ and $\hat{b}$ for all possible values of the black hole spin is plotted in Fig. 3. The coefficient $\hat{a}$ has a finite value in the extremal black hole limit $a\rightarrow\frac{1}{2}\big|1-x_{\alpha}\big|$, while $\hat{b}$ diverges.

The analysis of the integrand function $\omega(\phi)$ leads to $\omega<1$ for $a>0$ and to $\omega>1$ for $a<0$. In the particular case of Schwarzschild black hole when $a=x_{\alpha}=0$, $\omega=1$. In the case when $\omega<1$, $\overline{\phi}_{f}<\phi_{f}$ and after each loop the direct photons lag with $\Delta\phi$ towards the initial declination $\psi$. When $\omega>1$, $\overline{\phi}_{f}>\phi_{f}$ and the retro photons reach the same initial declination as advances with $\Delta\phi$ before completing a loop. For Schwarzschild black hole a precession is not present.

\subsection{Lensing at small declinations}

When we want to study the caustic structures and to compute the magnification of the images for small displacement from the equatorial plane we can write the polar lens equation in addition to the equatorial lens equation. In order to consider the lens geometry within the small declination hypothesis, we can set the source at height $h_{S}$ on the equatorial plane. We are setting the observer on height $h_{O}$ and during the whole time we will assume that the following relations
\begin{eqnarray}
  u\ll(h_{O},h_{S})\ll(D_{OL},D_{LS}) \label{HierDis} \end{eqnarray}
hold. After neglecting the higher order terms, using (\ref{HierDis}) we can build the polar lens equation, which connects the positions of the source and the observer \cite{BozQuazi}
\begin{eqnarray}
  h_{S} &=& h_{O}\left[ \frac{D_{OL}}{\overline{u}}\sin\overline{\phi}_{f}-\cos\overline{\phi}_{f} \right]- \nonumber \\
  &&\psi_{0}\left[ (D_{OL}+D_{LS})\cos\overline{\phi}_{f} - \frac{D_{OL}D_{LS}}{\overline{u}}\sin\overline{\phi}_{f} \right]. \label{PolEq}
\end{eqnarray}

Our purpose is to determine the inclination $\psi_{0}$ under which the observer sees the light ray. The solution of (\ref{PolEq}) for the n-th image is
\begin{eqnarray}
  \psi_{0,n} &=& \frac{(h_{S}+h_{O}\cos{\overline{\phi}_{f,n}})\overline{u}-h_{O}D_{LS} \sin{\overline{\phi}_{f,n}}}{D_{OL}D_{LS}\sin{\overline{\phi}_{f,n}}-\overline{u}(D_{OL}+D_{LS})\cos{\overline{\phi}_{f,n}}.  }\end{eqnarray}
The quantity $\overline{\phi}_{f,n}$ is the phase of the n-th image and may be obtained from Eq. (\ref{Phi_f}) once we know the solution $\theta_{n}$ of the equatorial lens equation.

According to the lensing geometry, in the polar direction we can write the parameter $h$ of the incoming light trajectory
as follows
\begin{eqnarray}
  h &=& h_{O}+D_{OL}\psi_{0}.\end{eqnarray}
In the approximation $\psi\ll1$ and $h\ll{u}$ the height $h_{n}$ of the n-th image is
\begin{eqnarray}
  h_{n} &=& \frac{\overline{u}(h_{S}D_{OL}-h_{O}D_{LS}\cos{\overline{\phi}_{f,n}})}{D_{OL}D_{LS}\sin{\overline{\phi}_{f,n}}-\overline{u}(D_{OL}+D_{LS})\cos{\overline{\phi}_{f,n}}}.\end{eqnarray}

Following (\ref{HierDis}) we can see that in the neighborhood of $\overline{\phi}_{f}=k\pi$ the denominators of the $\psi_{0,n}$ and $h_{n}$ vanish and both quantities diverge. Then equation
\begin{eqnarray}
 K(\gamma) &=& \overline{u}(D_{OL}+D_{LS})\cos{\overline{\phi}_{f}}-D_{OL}D_{LS}\sin{\overline{\phi}_{f}}=0 \label{CausticEq_K} \end{eqnarray}
defines the position of the caustic points. From Eq. (\ref{Phi_f}) for the phase $\overline{\phi}_{f}$ and the formula for the deflection angle (\ref{DefAng}) and with the help of the equatorial lens equation $\gamma=-\alpha(\theta_{n}^{0})$ mod $2\pi$ the caustic points equation for Kerr--Sen black hole takes the form
\begin{eqnarray}
 -\left(\frac{\gamma+\bar{b}}{\bar{a}}\right)\hat{a}+\hat{b} &=& k\pi. \end{eqnarray}
The solution
\begin{eqnarray}
 \gamma_{k} &=& -\bar{b}+\frac{\bar{a}}{\hat{a}}(\hat{b}-k\pi)\end{eqnarray}
shows that for each $k$ one caustic point for direct photons and one caustic point for retrograde photons exists. For $k=1$ we have weak field caustic points and an azimuthal shift about $\pi$ and strong field limit caustic point when $k\geq2$.
\begin{figure}
  \includegraphics[width=8cm]{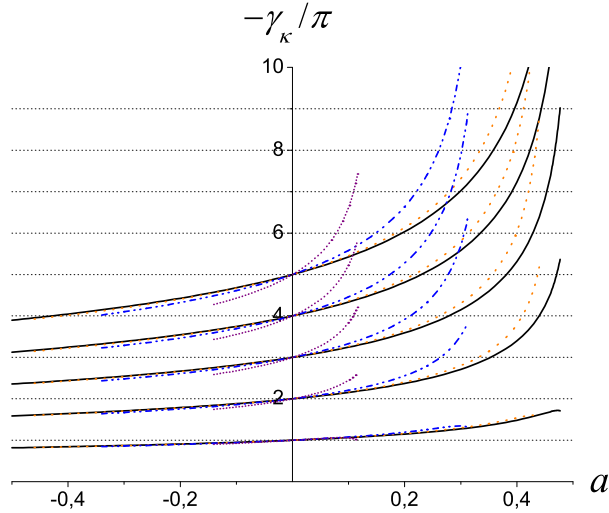}\\
  \caption{The angular position of the first five relativistic caustic points: $k=2$, 3, 4, 5, 6 from below to above as a function of the Kerr--Sen black hole angular momentum for $x_{\alpha}$ equal to 0 (black, solid lines), 0.08 , 0.32, 0.72 (purple, short, dotted lines).}\label{KS_Caustics}
\end{figure}

What is the magnification of the created images formed by the source close to the caustic points? The magnification, defined as the ratio of the angular area element of the image and the corresponding angular element of the source that the observer would see if there were no lens, in the considered lens geometry is
\begin{eqnarray}
 \mu &=& \frac{d^2\cal{A}_{I}}{d^2\cal{A}_{S}}=\frac{(D_{OL}+D_{LS})^2}{D_{LS}}\frac{1}{|J|}.\end{eqnarray}
From the lens map, which has the form
\begin{eqnarray}
  &&\gamma = \gamma(\theta), \\
  &&h_{S} = h_{S}(\theta, \psi_{0}),\end{eqnarray}
we can write the Jacobian determinant
\begin{eqnarray}
  |J| &=& \bigg|\frac{\partial\gamma}{\partial\theta}\frac{\partial{h}_{S}}{\partial\psi_{0}}\bigg|=\frac{\overline{u}u_{ps}e_{\gamma}}{\bar{a}D_{OL}|\overline{u}(D_{OL}+D_{LS})\cos{\overline{\phi}_{f}}-D_{OL}D_{LS}\sin{\overline{\phi}_{f}}|}, \label{JacobiDet}\end{eqnarray}
with
\begin{eqnarray}
e_{\gamma}=e^{\frac{\bar{b}+\gamma}{\bar{a}}},\end{eqnarray}
where $\gamma$ codes the number of loops made by the photon and take negative values, besides $\gamma$ mod $2\pi$ is the angular position of the source with respect to the lens position. $n=\frac{\pi-\gamma}{2\pi}$ is the number of loops done by the photon. In the particular case $\gamma=-2m\pi$, the source is situated behind the lens, while, when $\gamma=-(2m+1)\pi$, the source is situated before the lens.

In Fig. \ref{KS_Caustics} the positions of the first five relativistic caustic points are plotted as a function of the
Kerr--Sen black hole angular momentum for different values of $x_{\alpha}$. All caustic points coincide in the situation
$a=x_{\alpha}=0$ when the Schwarzschild metric is recovered. In this case $\gamma_{k}=-(k-1)\pi$. Then, the first
relativistic, caustic point of the source before the lens is $\gamma_{2}=-\pi$ and as follow $\gamma_{3}=-2\pi$ is the
caustic point of the source behind the lens. All caustic points are anticipated for negative $a$ and delayed for positive
$a$. When $x_{\alpha}$ grows the caustic curves become ever more shifted from one to other, while at large angular
momentum they move very far from the initial position. In the vicinity of an extremal black hole angular momentum, where the
strong deflection limit approximation fails the relativistic caustic points diverge. Therefore, the graphics are truncated
in the vicinity of the separable point where the divergence starts.

From Eq. (\ref{JacobiDet}) we can see that the Jacobi determinant diverges in the caustic points. Therefore, in order to write the magnification for the enhanced images created by a source close to the caustic points we can expand Eq. (\ref{CausticEq_K}) around $\gamma_{k}$
\begin{eqnarray}
 K(\gamma) &\simeq& K^{\prime}(\gamma_{k})(\gamma-\gamma_{k})=-\frac{\hat{a}D_{OL}D_{LS}}{\bar{a}}(\gamma-\gamma_{k}(a)).\end{eqnarray}
The formula for the magnification of the enhanced images then takes the form
\begin{eqnarray}
  &&\mu_{k}^{enh}=\frac{(D_{OL}+D_{LS})^2}{D_{OL}^{2}D_{LS}^{2}}\frac{\overline{\mu}_{k}(a)}{|\gamma-\gamma_{k}|},\\
  &&\overline{\mu}_{k}(a)=\frac{\overline{u}(a)u_{ps}(a)e_{\gamma_{k}(a)}}{\hat{a}},\\
  &&\,\,\,\,\,\,\,\,\,\,\,\,\,\,\,\,\,\,\,\, e_{\gamma_{k}(a)}=e^{ \frac{\hat{b}-k\pi}{\hat{a}}}. \label{exp_gamma_k}\end{eqnarray}
The quantity $\overline{\mu}_{k}$ in essence represents the magnifying power of the Kerr--Sen black hole close to caustic points.

\begin{figure}
\begin{minipage}[c]{.50\textwidth}
\includegraphics[width=8cm]{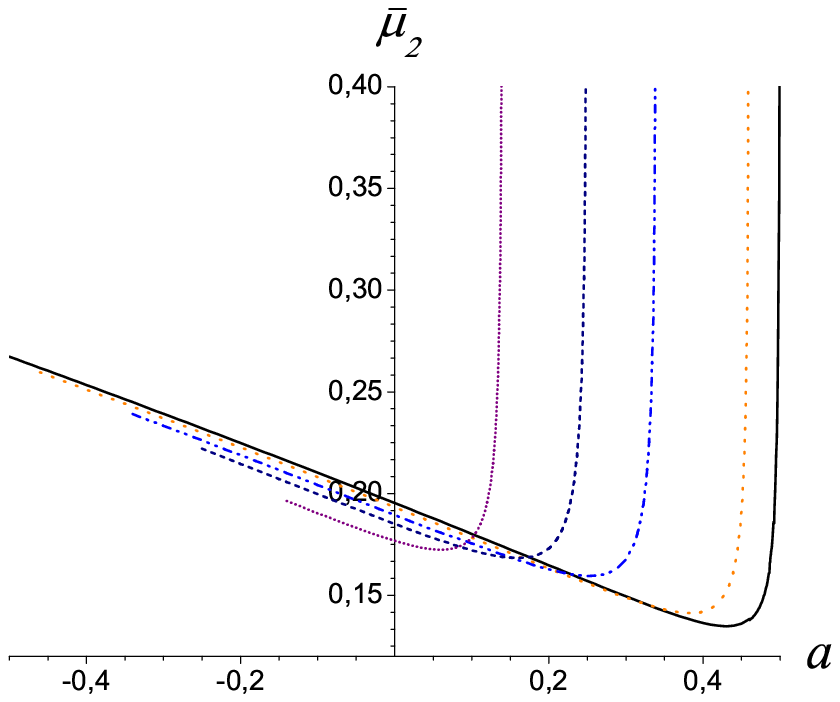}
\end{minipage}%
\begin{minipage}[c]{.50\textwidth}
\includegraphics[width=8cm]{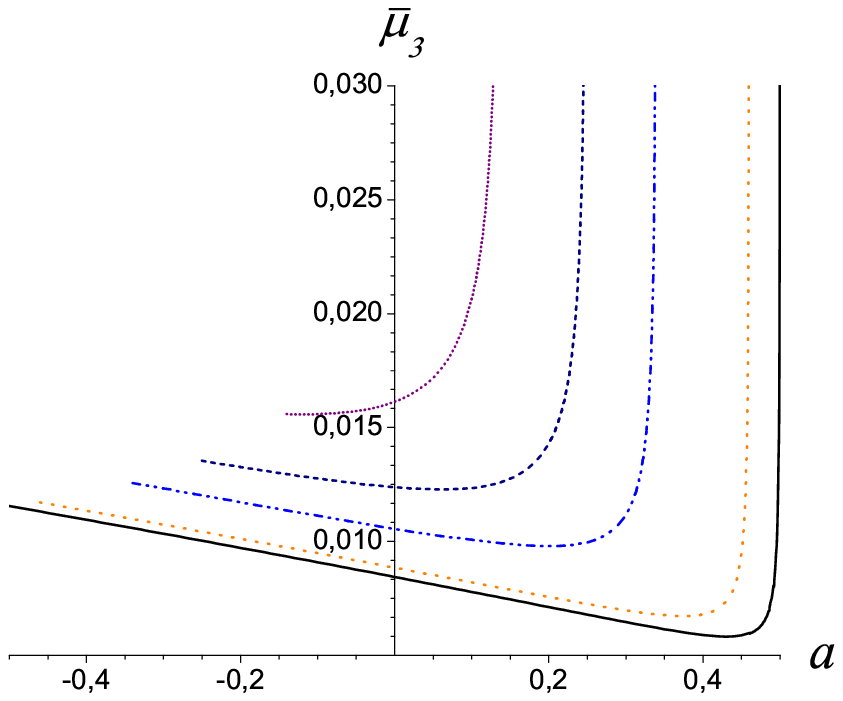}
\end{minipage}
\begin{minipage}[c]{.50\textwidth}
\vspace{0.5cm}
\includegraphics[width=8cm]{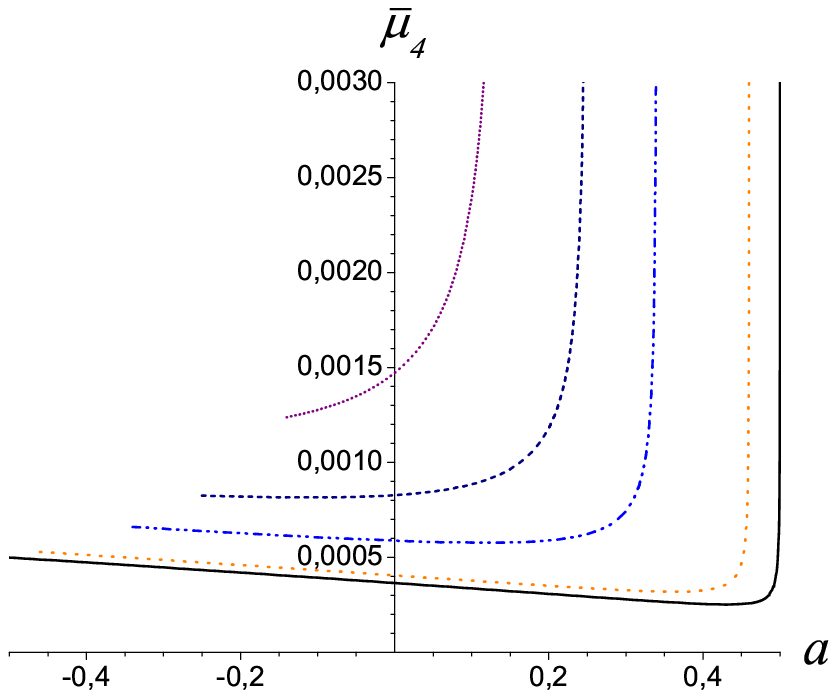}
\end{minipage}%
\begin{minipage}[c]{.50\textwidth}
\includegraphics[width=8cm]{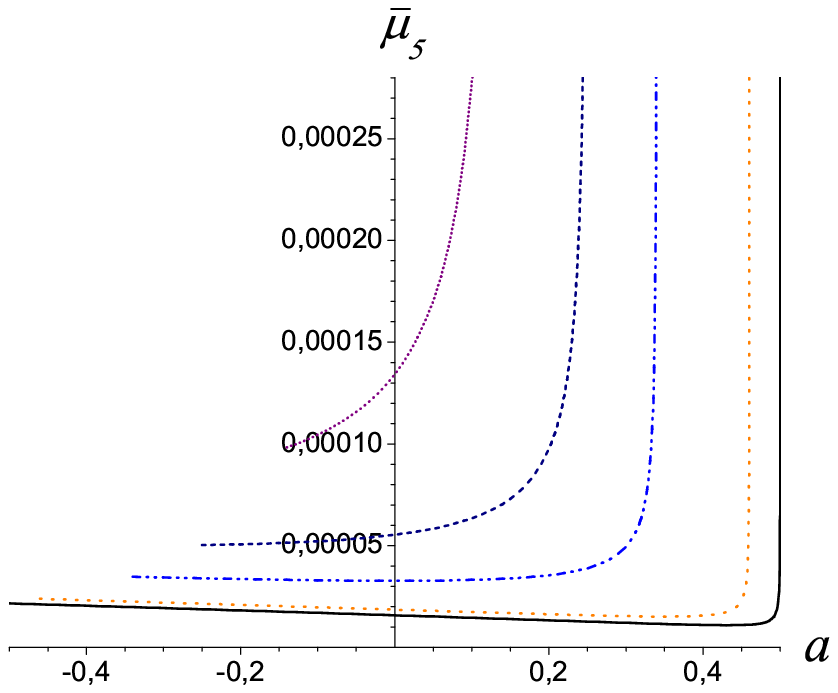}
\end{minipage}
\begin{minipage}[c]{.50\textwidth}
\vspace{0.5cm}
\includegraphics[width=8cm]{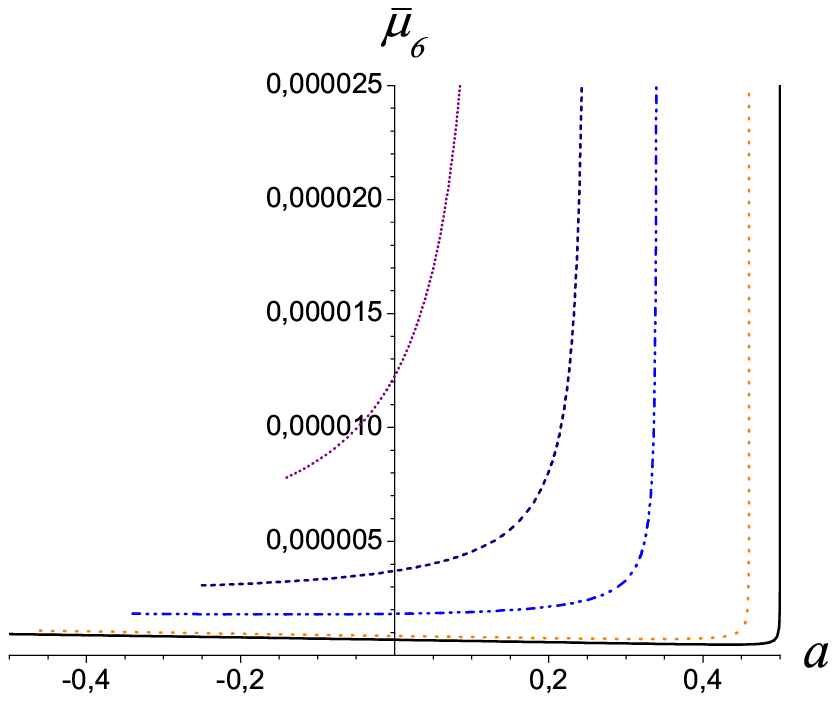}
\end{minipage}
    \caption{The magnifying power at the caustic point $\gamma_{2}$, $\gamma_{3}$, $\gamma_{4}$, $\gamma_{5}$, $\gamma_{6}$ as a function of Kerr--Sen black hole angular momentum for $x_{\alpha}$ equal to 0 (black, solid line), 0.08 , 0.32, 0.5, 0.72 (purple, short, dotted line) from below to above.}\label{KS_Magnifications}
\end{figure}

In Fig. \ref{KS_Magnifications} the numerically evaluated magnifying powers $\overline{\mu}_{2}$ to $\overline{\mu}_{6}$ at the caustics points from $\gamma_{2}$ to $\gamma_{6}$ respectively are plotted. The magnification decreases with growth of $|a|$ linearly for smaller values of $x_{\alpha}$, only grows for bigger plotted values of $x_{\alpha}$ and diverges for all plotted values of $x_{\alpha}$ when $a$ approaches the extremal value $\frac{1}{2}|1-x_{\alpha}|$. The magnifications of the neighbouring enhanced images falls rapidly with the winding of the photons increasing ever more with $x_{\alpha}$ and diverges at the extremal value of the black holes angular momentum, where the strong deflection limit approximation fails.
\begin{figure}
  \includegraphics[width=8cm]{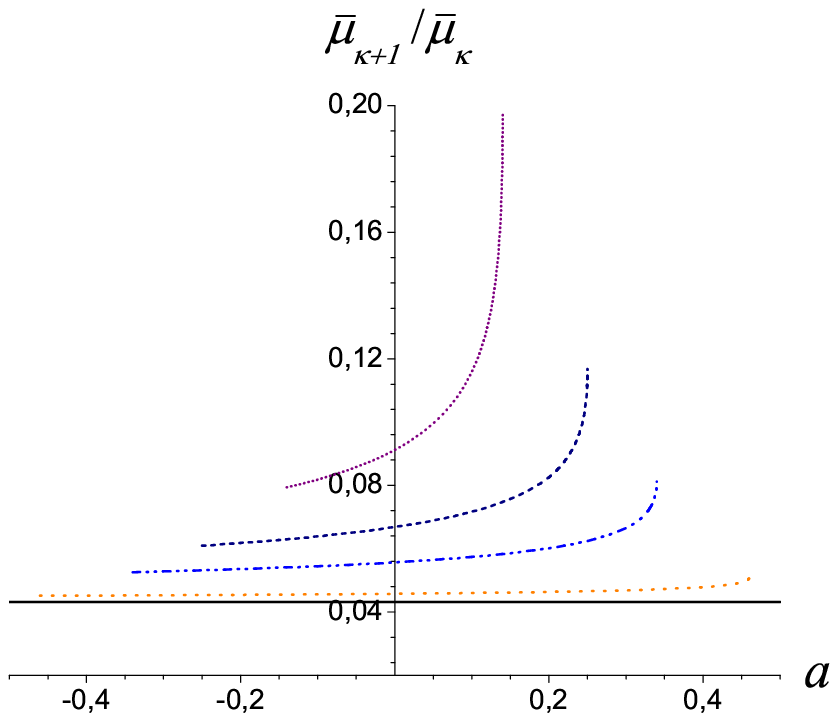}\\
  \caption{The ratio $\frac{\overline{\mu}_{k+1}}{\overline{\mu}_{k}}$ of the magnifying power at a neighbouring caustic points as a function of the Kerr--Sen black hole angular momentum for $x_{\alpha}$ equal to 0 (black, solid line), 0.08 , 0.32, 0.72 (purple, short, dot line) from below to above.}\label{KS_DeltaMagn}
\end{figure}

The shapes of $\overline{\mu}_{k}$ are identical for every $k$, but with factor (\ref{exp_gamma_k}) they are distinguished as follows
\begin{eqnarray}
 \frac{\overline{\mu}_{k+1}}{\overline{\mu}_{k}} &\simeq& e^{-\frac{\pi}{\hat{a}}}. \end{eqnarray}
For a Kerr--Sen black hole the ratio $\frac{\overline{\mu}_{k+1}}{\overline{\mu}_{k}}$ depends on $x_{\alpha}$ and the angular momentum, and is plotted in Fig. \ref{KS_DeltaMagn}.

\newpage

\section{CRITICAL CURVES AND CAUSTIC STRUCTURE}

In the basic case of Schwarzschild black hole when the light source, the lens and the observer are strictly aligned
we have a large weak field Einstein ring and an infinite series of concentric relativistic Einstein rings, very close to
the minimum impact angle $\theta_{\infty}$ \cite{Boz2}.

\begin{figure}
\includegraphics[width=8cm]{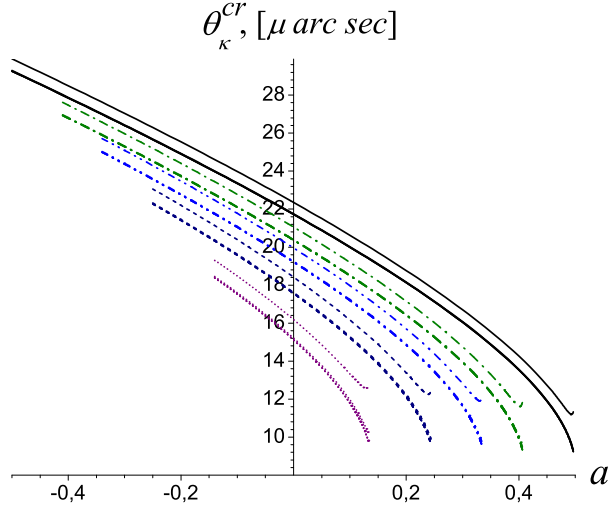}
\caption{Intersections of the strong deflection limit critical curves with the equatorial plane for $k$ from 2 to 6 (from above to below) as a function of a Kerr--Sen black hole angular momentum for $x_{\alpha}$ equal to 0 (black, solid lines), 0.18, 0.32, 0.5, 0.72 (purple, short, dotted lines). With an accuracy of the order of 0.01 $\mu$ $arc$ $sec$ the critical curve points $\theta_{k}^{cr}$ coincide with the positions of the innermost images $\theta_{\infty}$ for $k\geq4$.}\label{Theta_cr_KS}
\end{figure}

For the Kerr--Sen metric in the quasi-equatorial approximation in the strong deflection limit we can calculate the points of the intersections
of the critical curve with the equatorial plane, which are
\begin{eqnarray}
   &&\theta_{k}^{cr} \simeq   \theta_{k}^{0,cr}\left(1-\frac{u_{ps}e_{\gamma_{k}}(D_{OL}+D_{LS})}{\bar{a}D_{OL}D_{LS}}\right),\\
  &&\theta_{k}^{0,cr} = \frac{u_{ps}}{D_{OL}}(1+e_{\gamma_{k}}).\end{eqnarray}
The numerical results are plotted in the Fig. \ref{Theta_cr_KS} for the black hole in the center of our Galaxy according to the current estimates for the black hole mass \cite{GenSch} point $M_{BH}=3.6\times10^{6}M_{\odot}$ and assuming that $D_{OL}=8.5$ kpc and $D_{LS}=1$ kpc. For every value of $x_{\alpha}$ we have a family of characteristics plotted for $k$ from 2 to 6, where for $k$ from 3 to 6 the curves become very close to each other. The critical curves diverge at the extremal value of the angular momentum, when the strong deflection limit approximation breaks down. Close to the point of the divergence the graphics are not shown. For all values of $x_{\alpha}$ the critical points are closer to the optical axis for positive $a$ in case of direct photons and further from the optical axis for the negative $a$ when we consider opposite photons. For the static case when the GMGHS black hole is realized the critical points are not shifted. Therefore, as seen from the side of the observer, the critical curves are distorted and shifted towards the western side, if north is the direction of the spin. With an accuracy of the order of 0.01 $\mu$ $arc$ $sec$ the critical curve points $\theta_{k}^{cr}$ coincide with the position of the innermost images $\theta_{\infty}$ for $k\geq4$. From this point of view Fig. \ref{Theta_cr_KS} also shows the positions of the outermost relativistic images $\theta_{\infty}$.

\begin{figure}
\begin{center}
\includegraphics[width=9.6cm]{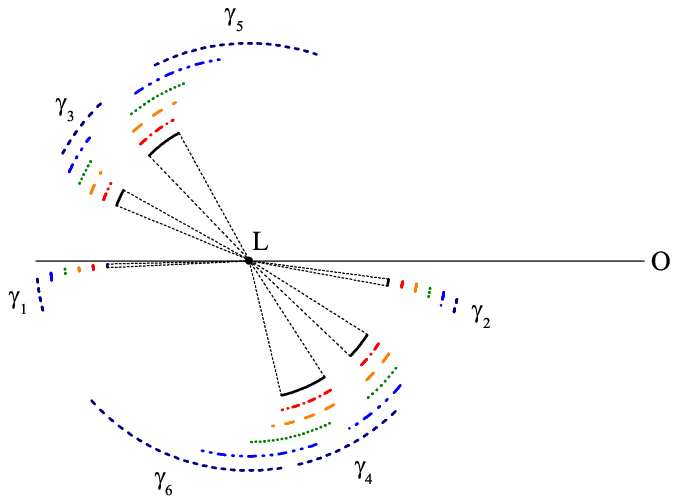}
\caption{The first six caustics of the Kerr--Sen black hole lens for $a=0.1$, marked by arcs between $\gamma_{k}(-|a|)$ and $\gamma_{k}(|a|)$ for $x_{\alpha}$ equal to 0 (black, solid lines), 0.02, 0.08, 0.18, 0.32, 0.5 (navy, short, dashed lines).}\label{Cusps_KS_1}
\includegraphics[width=9.6cm]{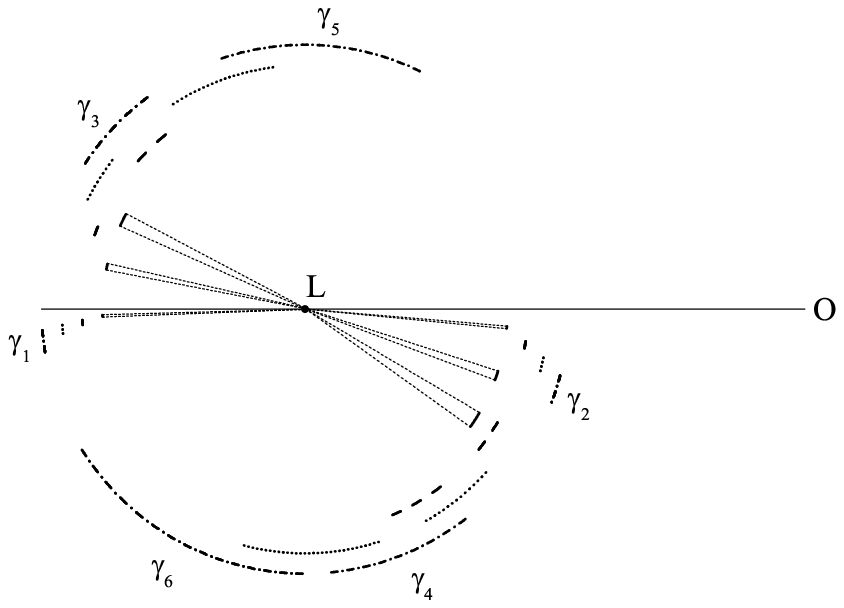}
\caption{The first six caustics of the Kerr--Sen black hole lens for $x_{\alpha}=0$, marked by arcs between $\gamma_{k}(-|a|)$ and $\gamma_{k}(|a|)$ for angular momentum $a$ equal to 0.05 (black, solid lines), 0.09, 0.13, 0.17 (black, short, dashed, dotted lines).}\label{Cusps_KS_2}
\end{center}
\end{figure}

\begin{figure}
\begin{center}
\includegraphics[width=9.6cm]{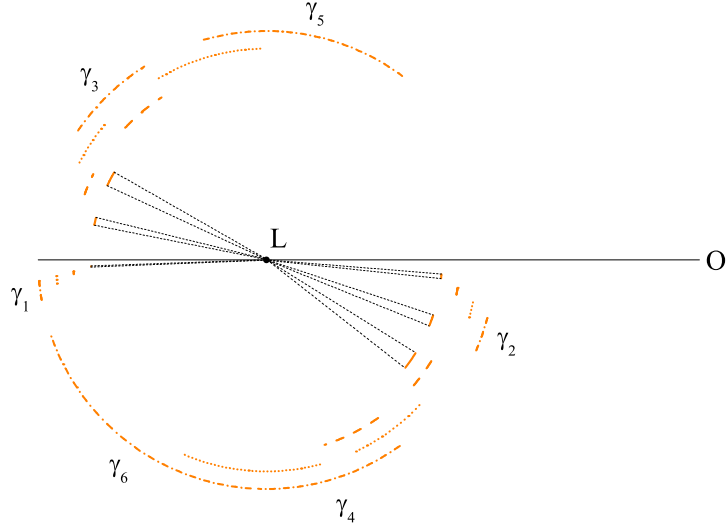}
\caption{The first six caustics of the Kerr--Sen black hole lens for $x_{\alpha}=0.08$, marked by arcs between $\gamma_{k}(-|a|)$ and $\gamma_{k}(|a|)$ for angular momentum $a$ equal to 0.05 (orange, solid lines), 0.09, 0.13, 0.17 (orange, short, dashed, dotted lines).}\label{Cusps_KS_3}
\end{center}
\end{figure}

$\gamma_{k}(-|a|)$ and $\gamma_{k}(|a|)$ represents the intersections of the $k$-th caustics with the equatorial plane.
For the static case ($a=0$) of GMGHS black hole the caustic points coincide with the caustic points for the Schwardzschild
black hole and they are distributed over the optical axis as follows $\gamma_{k}=-(k-1)\pi$. For $a\neq0$ the relativistic
caustics $\gamma_{2}$, $\gamma_{4}$, $\gamma_{6}$ are shifted toward the western side, while the relativistic caustic points $\gamma_{3}$, $\gamma_{5}$ are shifted towards the eastern side. The projections of the caustic curves on the
equatorial plane are plotted in Fig. \ref{Cusps_KS_1}, \ref{Cusps_KS_2}, \ref{Cusps_KS_3}, as seen from the north direction
in order to disentangle the influence of the two parameters $x_{\alpha}$ and $a$ over the caustics. Generally the
behaviour of the caustics as a
function of the black hole angular momentum and the parameter $x_{\alpha}$ are the same.
For all values of $x_{\alpha}$ and the black hole angular momentum $a$ the relativistic caustics drift in the clockwise
direction, while the non-relativistic caustics $\gamma_{1}$ stay close to the optical axis only for small value of
$x_{\alpha}$ and $a$, and drift in counterclockwise direction with the increase of $x_{\alpha}$ and $a$. With the
increase of $k$ the caustics become larger and further in comparison to their initial position with respect to optical
axis. Moreover, as $x_{\alpha}$ and $a$ grow, the caustics drift in clockwise direction with respect to the caustics for
smaller $x_{\alpha}$ and $a$. From the numerical results in the Fig. \ref{KS_Caustics} we can see that at height angular
momentum and $x_{\alpha}$ the caustics may become enormous. There the strong deflection limit fails and we have not
considered this situation.

\section{DISCUSSIONS AND SUMMARY}

In this paper we discuss the features of light propagation close to the equatorial plane of the rotating dilaton-axion black hole. Modelling the massive compact object at the center of the galaxy as a Kerr--Sen black hole, we estimate the numerical values of different strong-lensing parameters. When compared to the corresponding lensing observables due to the Kerr black hole, it is found that there is a significant dilaton-axion effect present on the observable parameters.

As the Kerr--Sen solution of the field equation describes a charged, stationary, axially-symmetric black hole we can obtain all lensing phenomena for the GMGHS charged, static spherically-symmetric black hole when the angular momentum $a=0$ and for the Schwarzschild black hole when $a=x_{\alpha}=0$. A generic source not aligned to the optical axis gives rise to extremely faint relativistic images, while a point source aligned to the optical axis produces infinitely bright images. For static--spherically symmetric black holes the caustics are points which are disposed over the optical axis, therefore a source close to one caustic is simultaneously close to all other caustics and gives rise only to enhanced images.

In the case $a\neq0$ and $x_{\alpha}=0$, when the Kerr lensing is recovered, the caustics drift away from the optical axis and only one image of all other composed infinite series of images can be enhanced, while the rest of the images will be quite faint. If we set the source on the caustic point $\gamma_{k}(|a|)$ then the outermost relativistic image on the western side will be enhanced, while if the source is at $\gamma_{k}(-|a|)$, then the enhanced image will be the first on the eastern side. For the source disposition at the caustic points $\gamma_{k+1}(|a|)$ or $\gamma_{k+1}(-|a|)$ the enhanced relativistic image will be the second image respectively on the western or the eastern side. If the source is inside the caustic, the parity of the corresponding image changes parity and two additional images appear with the same parity as the original image. The created images move rapidly in vertical direction and they are not liable to description through the quasi-equatorial approximation \cite{Boz2,Rhie}.

In Kerr--Sen lensing, when we have $a\neq0$, $x_{\alpha}\neq0$ the situation reminds of the Kerr lensing with several  correction. First of all, the equatorial critical points $\theta_{k}^{cr}$ decrease with the increase of $x_{\alpha}$ for the whole spectrum of $a$. The second consequence is that a correction of the position of relativistic caustic points exist in the case $x_{\alpha}\neq0$, which influences the caustic lines $\gamma_{k}$, as they become longer with the increases of $x_{\alpha}$ and black hole angular momentum $a$. The third consequence is that the magnifying power $\mu_{k}$ for the $k$-th caustic point, becomes larger with the increase of $x_{\alpha}$ for $k>2$, unlike the case $k=2$, where the magnifying power decreases with the increase of $x_{\alpha}$. Moreover for $k=2$ (the first strong caustic), the magnifying power decreases with the increase of $a$ for every $x_{\alpha}$, while from $k=3$ to $k=6$, a value $x_{\alpha}^{'}$ exists at which $\bar{\mu}_{k}$ does not depend on the angular momentum. For all values $x_{\alpha}>x_{\alpha}^{'}$, the magnifying power only grows. The fourth consequence is that, the ratio $\frac{\overline{\mu}_{k+1}}{\overline{\mu}_{k}}$ of the magnifying power at neighbouring caustic points depends on the angular momentum when $x_{\alpha}\neq0$ and grows with $x_{\alpha}$ as the function $e^{-\frac{\pi}{\hat{a}}}$.

The caustics become longer with the increases of $x_{\alpha}$ and/or the black hole momentum $a$. They also drift away from the optical axis and shift in clockwise direction in comparison to the Kerr caustics. Through $a=0$ the source is simultaneously close to all caustics and gives rise only to the enhanced images, while for the Kerr--Sen black hole the source can be close to only one caustic at the same time and thus produces only one enhanced image. Photons which move in the equatorial plane in direction opposite to the direction of rotation of the black hole have a bigger impact parameter compared to the impact parameter of the direct photons, which means that the retrograde photons are captured more easily by the black hole than the direct photons. The probability for absorbtion for the two kinds of photons decreases when the parameter $x_{\alpha}$ grows. The magnification of the image strongly depends on the charge of the black hole and increases with $x_{\alpha}$.

When comparing these results to the corresponding observable lensing quantities due to the Schwarzschild and Kerr black holes, it is found that there is a significant dilaton-axion field effect present in the observable parameters. This means that the mentioned black holes solutions are quite different with respect to the observational effects in terms of the strong deflection limit. Hence, detecting the relativistic images, which might be possible in the near future through the very long baseline interferometry high resolution imaging, we will be able to either accommodate or rule out the Kerr--Sen black hole candidate as a possible lensing object.

\begin{acknowledgments}
We wish to express our gratitude to K. S. Virbhadra and V. Bozza for the helpful correspondence. We are also grateful to Ivan Stefanov for reading the manuscript. This work was partially supported by the Bulgarian National Science Fund under Grants MUF04/05 (MU 408), VYF-14/2006 and the Sofia University Research Fund N.60.
\end{acknowledgments}

\end{document}